\documentclass[aps,pra,10pt,amsmath,amssymb,A4,twocolumn,superscriptaddress,showpacs,floatfix]{revtex4-1}
\usepackage{bm}
\usepackage{epsfig}
\usepackage[usenames]{color}
\usepackage{soul}
\usepackage{graphicx}
\usepackage[hypertex]{hyperref}
\usepackage{srctex}

\begin{document}

\title{On the Topological Features of the Helical Phase Transition in MnSi.}
\author{S.~M.~Stishov}
\email{stishovsm@lebedev.ru}
\affiliation{P. N. Lebedev Physical Institute, Leninsky pr., 53, 119991 Moscow, Russia}
\author{A.~E.~Petrova}
\affiliation{P. N. Lebedev Physical Institute, Leninsky pr., 53, 119991 Moscow, Russia}
\author{A. M. Belemuk}
\affiliation{Institute for High Pressure Physics of RAS, 108840 Troitsk, Moscow, Russia}

\begin{abstract}
Many decades of study have revealed very unusual properties of the helical phase transition in MnSi.  This situation is briefly described and illustrated in the present note. As one will be able to see, one peculiarity is that the phase transition point in MnSi is accompanied by extremes of different thermodynamic and kinetic quantities on the high temperature side of the transition, which look similar to a property of 2D systems. The whole situation can be tentatively described as a phase transformation of the helical phase of MnSi to the paramagnetic state, which occurs in two steps, first as a first order phase transition and then following a breakdown of topological objects such as spin vortices. 

\end{abstract}

\maketitle

\section{Introduction}

MnSi is a weak itinerant ferromagnetic 3d metal crystallizing in a non-centrosymmetric cubic B 20 structure with space group $P2_{1}3$, which contains no center of symmetry~\cite{stish,stish2}.
The magneto-ordered state in MnSi was found in~\cite{will}. The same work indicated that magnetic transformation in MnSi takes place at a temperature $\approx 29 K$. Helical magnetic order in MnSi was established in~\cite{ish}.
The magnetic structure of MnSi in zero magnetic field can be described as a system of planes parallel to the crystallographic plane (111) containing ferromagnetically ordered spins. The magnetic moment of each layer is rotated by a small angle relative to the magnetic moment of adjacent layers due to the Dzyaloshinsky-Moria (DM) interaction, which is different from zero in the case of non-centrosymmetric structures~\cite{dzy,mor}.  As a result, in the magnetically ordered phase the spins form an incommensurate helix with a wave vector ~ 0.036 \AA$^{-1}$ (corresponding to a period of 180\AA) in the [111] direction.
Figs.~\ref{Fig1} - \ref{Fig4} show the heat capacity, sound attenuation, coefficient of thermal expansion and elastic modulus $c_{11}$ of MnSi in the phase transition region along with corresponding data for KDP (KH$_{2}$PO$_{4}$) ferroelectric crystal. The KDP crystal is one of the model materials exhibiting a small first order phase transition close to a second order phase transition. A drastic difference in the character of the phase transitions in both cases is quite obvious. In contrast to the KDP case, the phase transition in MnSi is accompanied by extremes or shoulders of different thermodynamic and kinetic quantities. It is worth noting that the heat capacity maxima observed in MnSi above the phase transition point are surprisingly similar to a property of 2D systems.
\section{Heat Capacity}

\begin{figure}[htb]
\includegraphics[width=80mm]{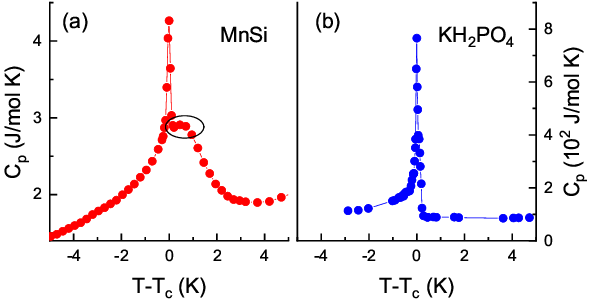}
\caption{\label{Fig1} 1 Heat capacity of MnSi at magnetic phase transition (a)~\cite{stish3,stish4}. Heat capacity of KDP at ferroelectric phase transition (b)~\cite{str}. The bump marked by the ellipse in (a) shows the peculiarity of the phase transition in MnSi when compared with other typical first-order phase transitions close to the second-order (see b).}
\end{figure}

\begin{figure}[htb]
\includegraphics[width=60mm]{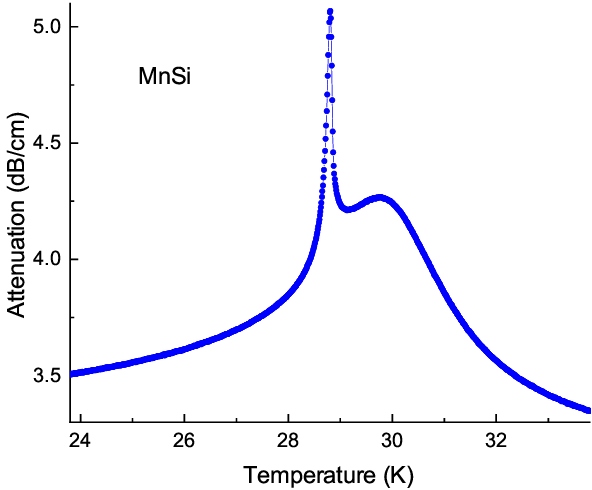}
\caption{\label{Fig2} Attenuation of longitudinal sound waves in MnSi shows non-trivial side maximum ~\cite{pet2} .}
\end{figure}

The first measurements of the heat capacity of MnSi on a sample of low quality were carried out by Fossett et al.~\cite{faw}.  Nevertheless, by extrapolating to zero $C_{p}/T$, Fawsett obtained a value for the electronic heat capacity coefficient $\gamma=0.038 J/mol K^{2}$ which is correct by modern standards.
However, an important feature (sharp peak) on the heat capacity curve of MnSi in the phase transition region was not found. Later, C. Pfleiderer found a heat capacity peak accompanied by a side maximum when measuring the heat capacity in MnSi~\cite{pfl}. However, prolonged annealing of MnSi samples was found to significantly reduce the side maximum, leading to the conclusion that it was defective in nature.  In subsequent work, the complex structure of the heat capacity anomaly at the phase transition in MnSi was confirmed~\cite{lam}, but no proper conclusions were drawn.  One way or another, the idea of the phase transition in MnSi as a phase transition of the second kind was fixed in the public consciousness.
New measurements of heat capacity on high quality crystals in magnetic fields up to 4 Tesla were carried out in ~\cite{stish3, stish4}(Fig.~\ref{Fig1}).
As follows from Fig.~\ref{Fig1}a the heat capacity curve $C_{p}$ (T) at B=0 is characterized by a sharp peak at $T_{c}\approx 29 K$ corresponding to the phase transition and a distinct “shoulder” on its high temperature side, in contrast to the behavior of the heat capacity at the phase transition in KDP Fig.~\ref{Fig1}b ~\cite{str}.

\section{Thermal expansion}

\begin{figure}[htb]
\includegraphics[width=80mm]{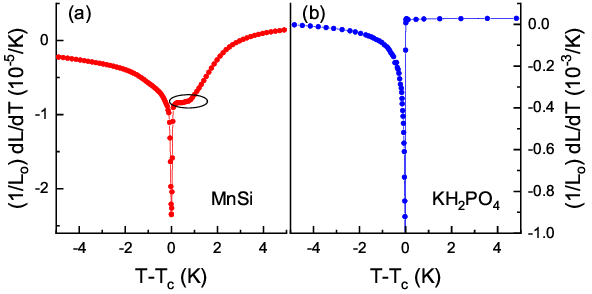}
\caption{\label{Fig3} Thermal expansion coefficient of MnSi at magnetic phase transition (a)~\cite{stish3,stish4}. Thermal expansion coefficient of KDP at ferroelectric phase transition near the tricritical point at pressure 2621.5 bar (b). Again, the shallow minimum indicated by the ellipse near the sharp negative peak in (a) is the peculiarity of the phase transition in MnSi compared to other typical first order phase transitions close to the second order ~\cite{zis} (see b).}
\end{figure}

For the first time, the thermal expansion of MnSi was measured in~\cite{faw}, where an extensive negative anomaly of the thermal expansion coefficient in the phase transition region was found. In~\cite{mat}, apparently also for the first time, a complex structure of the anomaly (noted earlier in the case of heat capacity) was observed, characterized, in addition to the main negative peak at the phase transition point, by a side maximum (shoulder) at a temperature a few tenths of a degree above $T_{c}$.
Detailed studies of the thermal expansion of MnSi have been carried out in~\cite{stish3,stish4}, the results of which are presented in Fig.~\ref{Fig3} and \ref{Fig8}. Fig.~\ref{Fig3}a shows the behavior of the linear thermal expansion coefficient of MnSi from data obtained using a capacitive dilatometer. These results are compared with the thermal expansion measurements of KDP~\cite{zis}. Fig.~\ref{Fig3}b.  Fig.~\ref{Fig8} clearly shows the global volume anomaly in MnSi during the magnetic phase transition~\cite{pet}.

\begin{figure}[h!]
\includegraphics[width=80mm]{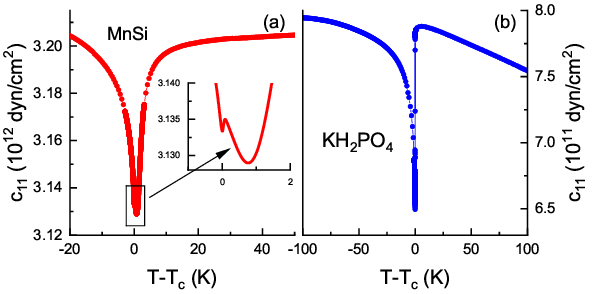}
\caption{\label{Fig4} Elastic module $c_{11}$ of MnSi  at magnetic phase transition (a)~\cite{pet2}.  The depicted vast minima of $c_{11}$  indicates that a tiny peak corresponding to the first order transition is only a minor feature of the global transformation in MnSi (In this connection see also Fig.~\ref{Fig3}). Nothing of the sort can be seen in Fig.~\ref{Fig4}b~\cite{pet3}.}
\end{figure}

\section{Elastic Properties}
The elastic properties and ultrasonic attenuation in MnSi were investigated using the digital pulsed ultrasonic technique in~\cite{pet2}. Some results of this work are shown in Figures~\ref{Fig2} and ~\ref{Fig4}a. Fig.~\ref{Fig4}b shows the contrasting behavior of the modulus of elasticity $C_{11}$ during the phase transition in KDP~\cite{pet3}. As can be seen from Fig.~\ref{Fig4}a, the magnetic phase transition in MnSi is accompanied by a deep anomaly (softening) of the elastic modulus $C_{11}$.  At the same time, the magnetic phase transition itself appears as a small jump of the elastic modulus localized on the low temperature side of the anomaly. Note that both of these features correlate perfectly with the behavior of the heat capacity, thermal expansion coefficient, and temperature coefficient of electrical resistivity~\cite{stish,stish2}, except that the sharp maxima and minima of these thermodynamic and transport properties at the phase transition point are replaced by modest jumps in the elastic moduli. This is exactly the behavior of the elastic moduli expected for a weak first order phase transition.
Let us now turn to Fig.~\ref{Fig2}, which illustrates the attenuation of ultrasonic waves. Curiously, the general structure of the attenuation curves (main peak and side maximum) is an almost perfect copy of the corresponding curves characterizing the behavior of heat capacity and coefficient of thermal expansion (with opposite signs). This result confirms earlier data~\cite{kus}.
The totality of the above data reveals a very important fact, which clearly shows that the magnetic phase transition at 28.8 K in MnSi is only one of the features of the global transformation accompanied by anomalies of physical quantities with extrema located at temperatures slightly higher than the phase transition temperature.
\section{Monte Carlo calculations}

\begin{figure}[htb]
\includegraphics[width=80mm,]{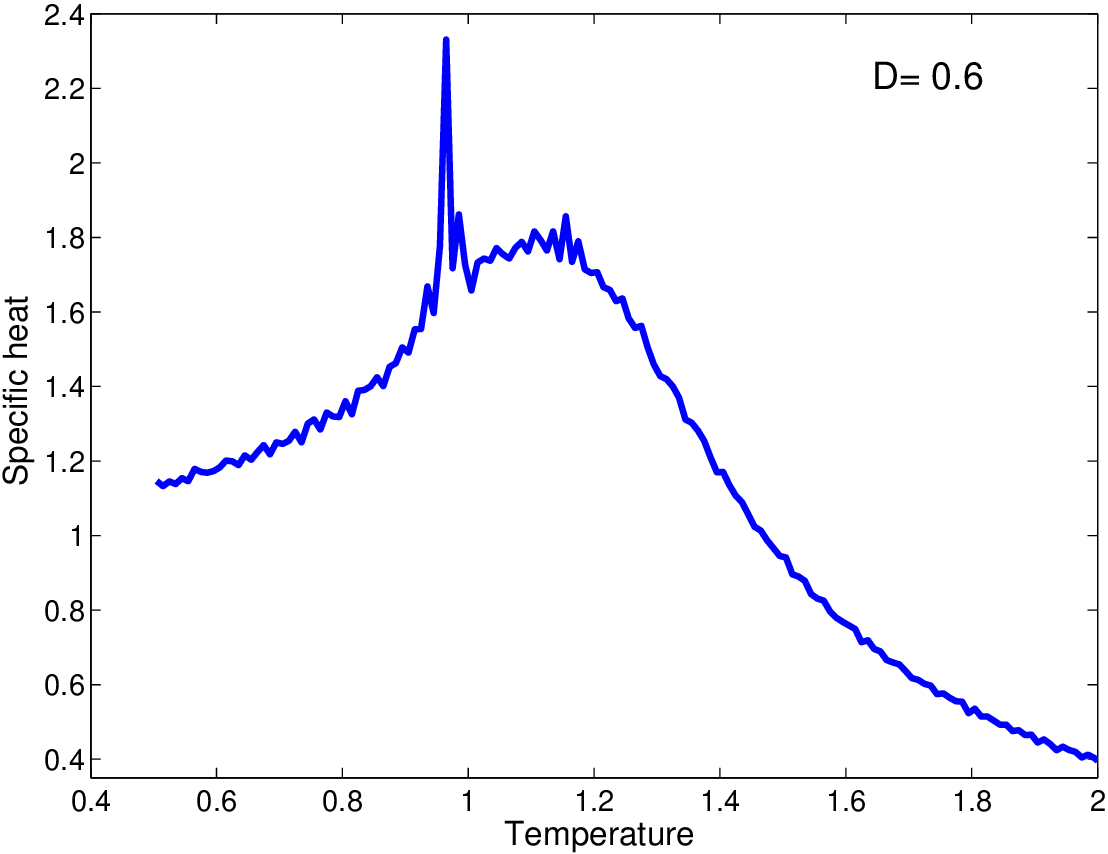}
\caption{\label{Fig5} Heat capacity at phase transition in 3D classical Heisenberg spin system with DM interaction by the Monte-Carlo calculations ~\cite{bel}. Coupling constants of exchange and DM interaction are J=1 and D=0.6. Temperature is given in units of constant $J= 1$. Phase transition point $T_{c} \cong 0.96$  .}
\end{figure}

\begin{figure}[h!]
\includegraphics[width=80mm]{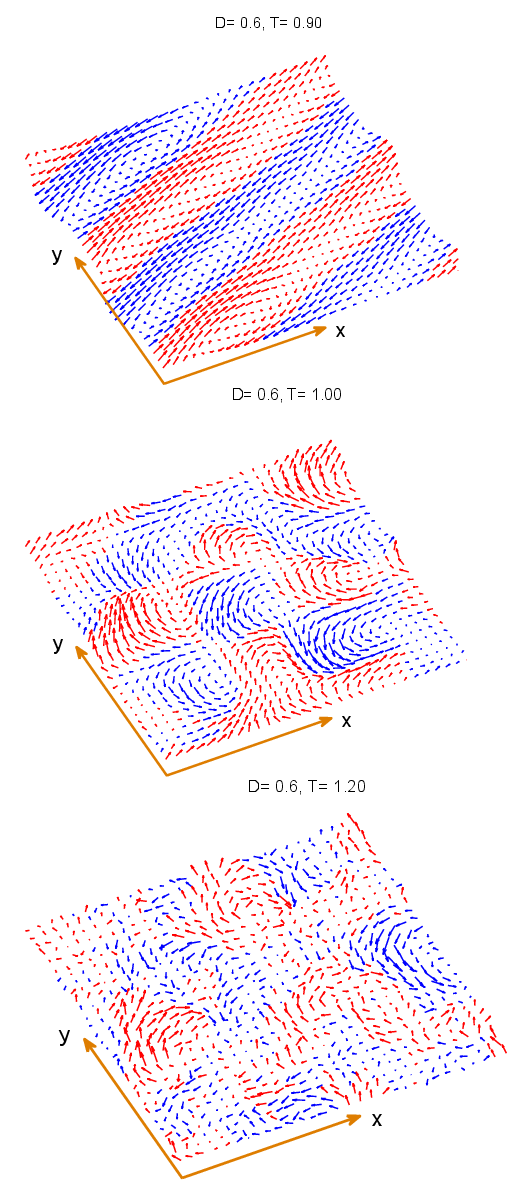}
\caption{\label{Fig6} Spin configurations at helical phase transition in classical chiral spin system Monte-Carlo calculations  at J=1 and D=0.6. Phase transition point $T_{c} \cong 0.96$. Vortex structure of spins is clearly visible at temperature above $T_{c}$, corresponding to side maximum (shoulder) in $C_{p}$~\cite{bel}   .}
\end{figure}

Monte Carlo simulations of the classical chiral spin system were performed in the papers~ \cite{Buhrandt,bel} and the heat capacity at the helical phase transition in the system is shown in Fig.~\ref{Fig5}. 
The spin configurations are shown in Fig.~\ref{Fig6} at different temperatures close to the magnetic transition. It can be seen that the spin configurations at temperatures above $T_{c} \cong 0.96$ have apparent vortex structure and correspond to the side maximum of $C_{p}$. These vortices are topological objects and it is not excluded that just their destruction gives the indicated feature in the heat capacity.
Note that in the neutron scattering studies~\cite{pap,grig} this feature is characterized as a strongly chiral fluctuation region. As we could see in the previous sections, the phase transition features in MnSi are quite unique and look like the 2D XY transition. What actually happens to the helical spin order in MnSi during heating? It was pointed out earlier that the helical order in MnSi can be described as a system of ferromagnet planes parallel to the (111) crystallographic planes, and each plane is rotated by a small angle relative to the neighboring planes by the Dzyloshinsky-Moria interaction. Thus, the disappearance of the helical order at the phase transition point implies the loss of the plain-plain correlation, which allows the spins to form structures with a shorter correlation length. The Monte Carlo data show that under certain conditions spin vortices can develop in MnSi with subsequent decay or destruction by temperature, which would lead to the thermodynamic anomalies. 
It is curious that the behavior of the heat capacity in the 3D Heisenberg model with DM interaction \cite{Buhrandt,bel} is somewhat analogous to that in the 2D XY system~\cite{tob,gup,Ngu} (see see discussion below).


\begin{figure}[h!]
\includegraphics[width=80mm,height=50mm]{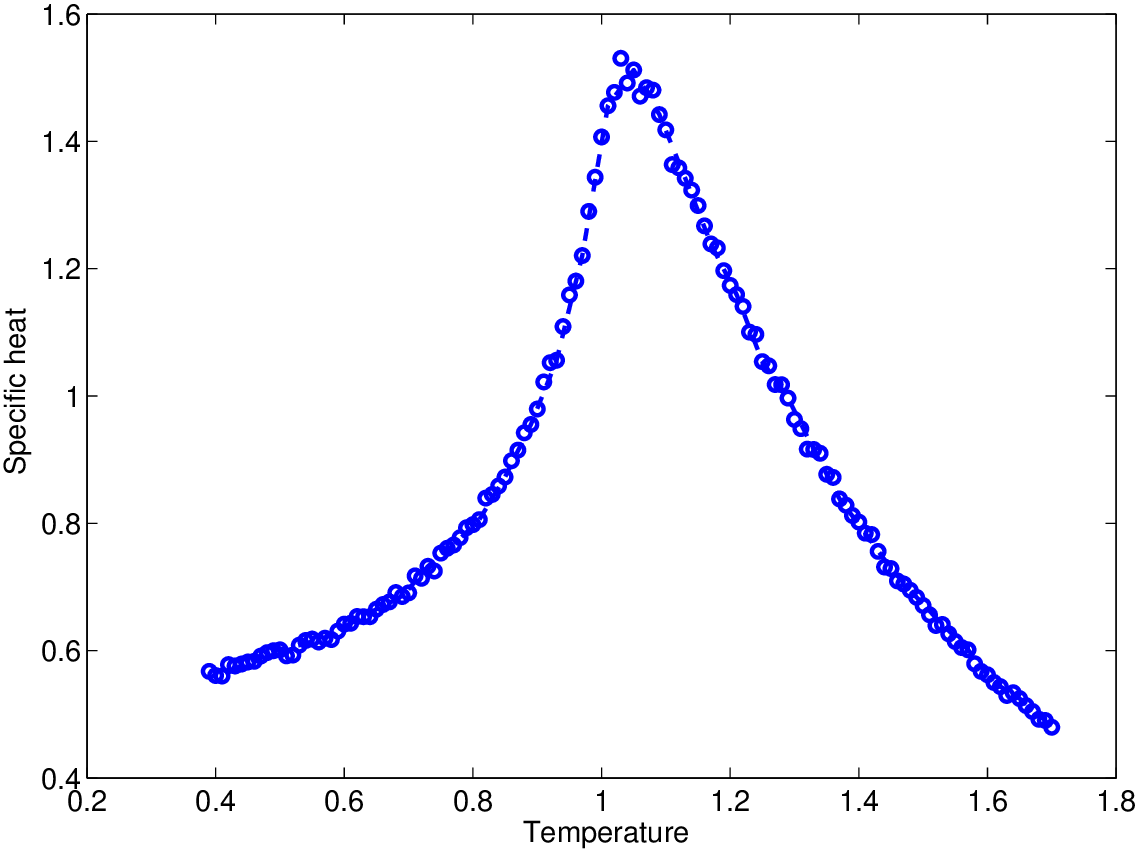}
\caption{\label{Fig7} Heat capacity in 2D XY classical spin system by the Monte-Carlo calculations with the  exchange interaction constant J=1. Compare it with the heat capacity plots in Fig.~\ref{Fig5} and \ref{Fig8}. Temperature is given in units of coupling constant $J= 1$, as in the 3D case The heat capacity has a smooth peak, which does not correspond to a phase transition.}
\end{figure}

\begin{figure}[h!]
\includegraphics[width=80mm]{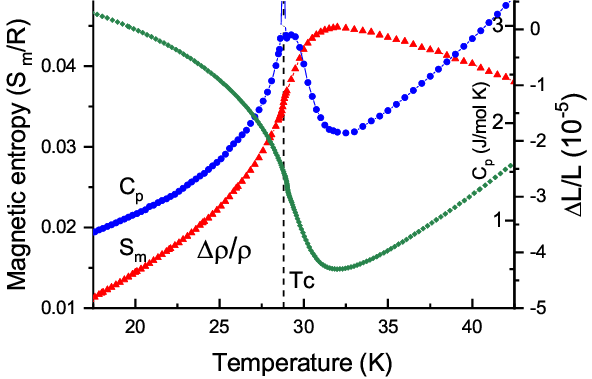}
\caption{\label{Fig8} Magnetic entropy, heat capacity $C_{p}$ and linear thermal expansion of MnSi in the vicinity of its phase transition. The peak in $C_{p}$ indicating first order phase transition is shown in a schematically for a better view of the rounded maximum. Modified from Ref. \cite{stish4}.The magnetic entropy was calculated by subtracting the electronic and phonon contributions~\cite{stish5}. }
\end{figure}

\section{Heat capacity in 2D XY model}
To see clearly a difference between 3D and  2D models we performed Monte Carlo calculations of the temperature behavior of the heat capacity for 2D XY model on a square lattice of linear size $L= 50$. The heat capacity is calculated from the fluctuations in the total energy $C(T)= (\langle E^2 \rangle- \langle E \rangle^2)/(L^2 T^2)$. We plot heat capacity per site obtained from a measurement of $25 \cdot 10^3$ spin configurations. We use standard Metropolis algorithm combined with over-relaxed sweeps \cite{gup}. Measurements are separated by $N_{or}= 20$ over-relaxed and $N_{met}= 20$ Metropolis sweeps.
Fig. \ref{Fig7} shows the behavior of the heat capacity in 2D XY system. The heat capacity demonstrates a smooth peak, which, as known~\cite{Kost}, does not correspond to any phase transition.

\section{Discussion}

As can be seen in the previous sections that in contrast to the KDP case, the phase transition at $\approx 29 K$ in MnSi is accompanied by extremes or shoulders of different thermodynamic and kinetic quantities. At the same time it is
obvious that the small helical first order phase transition in MnSi does not affect much the behavior of thermodynamic quantities with temperature variation. In fact, as seen in Fig.~\ref{Fig8}, a tiny entropy jump does not disturb the general behavior of entropy. The same is true for the elastic modulus and the thermal expansion  (Figs.~\ref{Fig4},\ref{Fig8}).  All this probably proves that the phase transition at 28 K, being a minor feature of the transformation, manifests only a loss of long-range helical order, but leaves  the general spin ability to form spiral structures practically intact (DM interaction still acting). Then, as the Monte Carlo data suggest, under certain conditions an interacting spin vortices can develop in MnSi above $T_{c}$ with subsequent decay or destruction by temperature, which would result in the thermodynamic response. Note that in the neutron scattering studies~\cite{pap,grig}  this site characterized as a strongly chiral fluctuation region. On the other hand, the magnetic entropy maximum as well as  minimuma in the heat capacity and thermal expansion in Fig.~\ref{Fig8} signify a final transformation of the spin system to the paramagnetic state.

\section{Conclusion}

We conclude that at temperatures slightly above $T_{c}$ topological spin excitations similar to those in the 2D XY model (see Fig.~\ref{Fig7}) occur in MnSi, which are responsible for the corresponding thermodynamic revelations. 

Finally, the whole situation can be tentatively described as a phase transformation of the helical phase of MnSi into the paramagnetic state, occurring in two steps,  first as a first order phase transition, then following by a breakdown of topological objects such as spin vortices. 
We're not sure yet if this breakdown can be called a topological phase transition.
	
\end{document}